\documentclass[prd,twocolumn,floatfix,amsmath,nofootinbib,amssymb,floatfix]{revtex4}
\usepackage{graphicx,color,dcolumn,booktabs,bm,multirow}
\usepackage{longtable,lscape}
\usepackage{txfonts}
\usepackage{overpic}
\usepackage{amssymb}
\usepackage{array}
\usepackage{indentfirst}
\usepackage{feynmf}   
\usepackage{slashed}  
\usepackage{cases}
\usepackage{color}
\usepackage{multirow}
\usepackage{epstopdf}
\usepackage{tabularx}
\usepackage{graphicx,color,dcolumn,booktabs,bm}
\usepackage[colorlinks,
citecolor=blue,
anchorcolor=red,
menucolor=red,
linkcolor=red,
filecolor=red,
runcolor=red,
urlcolor=blue,
frenchlinks=red]{hyperref}
\usepackage{braket}

\def\Zcs4000{$Z_{cs}(4000)^+$}

\begin{document}

\title{$Z_{cs}^+$ production in the $B^+$ decays process}
\author{Zhuo Yu$^{1}$}\email{zhuo@seu.edu.cn}
\author{Qi Wu$^{2}$}\email{wuqi@htu.edu.cn}
\author{Dian-Yong Chen$^{1,3}$\footnote{Corresponding author}}\email{chendy@seu.edu.cn}
\affiliation{$^1$ School of Physics, Southeast University, Nanjing 210094, China}
\affiliation{$^2$Institute of Particle and Nuclear Physics, Henan Normal University, Xinxiang 453007, China}
\affiliation{$^3$ Lanzhou Center for Theoretical Physics, Lanzhou University, Lanzhou 730000, China}
\date{\today}

\begin{abstract}
In the present work, we studied the $Z_{cs}^+$ exotic state production in the $B^+$ meson decays through the meson loop mechanism, where the $B^+$ meson decays into an anti-charmed meson and a charmed-strange meson pair, and consequently couples to $Z_{cs}$ and a light meson ($\phi,\rho^0,\omega,\eta,\eta^\prime$, and $\pi^0$) via exchanging a proper charmed or charmed-strange meson. Using the effective Lagrangian approach, we estimated the branching fractions of different channels find them to be on the order of $10^{-4}$ and their ratios are almost independent of the model parameter. The fit fractions of $Z_{cs}^+$ in different $B^+$ decay processes are also estimated, and we propose searching for $Z_{cs}^+$ in the $B^+ \to J/\psi K^+\eta^\prime$ process, which should be accessible to both Belle II and LHCb Collaborations.
\end{abstract}
\maketitle

\section{introduction}
\label{Sec:introduction}
The quark model categorizes the simplest hadrons as mesons, composed of a quark and an antiquark, and baryons, which consist of three quarks. However, since the first observation of $X(3872)$ by the Belle Collaboration in 2003~\cite{Belle:2003nnu}, an increasing number of exotic candidates have been reported by the LHCb, Belle, BaBar and BESIII Collaborations (see Refs.~\cite{Chen:2016qju,Hosaka:2016pey,Lebed:2016hpi,Esposito:2016noz,Guo:2017jvc,Ali:2017jda,Olsen:2017bmm,Karliner:2017qhf,Yuan:2018inv,Dong:2017gaw,Liu:2019zoy,Liu:2024uxn} for recent reviews). Among these states, the charmoniumlike states consisting of a $c\bar{c}$ pair have formed a large family. Unlike the neutral $X(3872)$, the first charged charmoniumlike state, $Z_c^-(4430)$, was observed in the $\pi^-\psi(2S)$ mass spectrum of the $B \to K \pi^- \psi(2S)$ process by the Belle Collaboration in 2007~\cite{Belle:2007hrb,Belle:2009lvn,Belle:2013shl}, and later confirmed by the LHCb collaboration in the same process~\cite{LHCb:2014zfx}. Following $Z_c^-(4430)$, several more charged charmonium-like states have been observed, including  $Z_c^-(4240)$, observed in the $\pi^- \psi^\prime$ invariant mass spectrum of the $B^0\to\psi^\prime \pi^-K^+$ decay~\cite{LHCb:2014zfx}, $Z_c^+(4050)$ and $Z_c^+(4250)$, seen in the $\pi^+ \chi_{c1} $ invariant mass spectrum of the $\bar{B}^0 \to K^- \pi^+ \chi_{c0}$ decay~\cite{Belle:2008qeq}, $Z_c^{\pm}(3900)$, reported in the $\pi^{\pm} J/\psi$ invariant mass spectrum of the process $e^+e^- \to \pi^+ \pi^- J/\psi$~\cite{BESIII:2013ris,Belle:2013yex}, $Z_c^+(4200)$, found in the $J/\psi \pi^+$ invariant mass spectrum of the $\bar{B}^0 \to J/\psi K^- \pi^+$ decay~\cite{Belle:2014nuw}, $ Z_c^-(4100)$, observed in $\eta_c \pi^-$ invariant mass spectrum of the $B^0 \to \eta_c(1S)K^+ \pi^-$ decay~\cite{LHCb:2018oeg}, and $Z_c^{\pm}(4020)$ reported in the $\pi^\pm h_c$ invariant mass spectrum of the process $e^+ e^- \to \pi^+ \pi^- h_c$~\cite{BESIII:2013ouc}.

Among these $Z_c$ states, $Z_c^{\pm}(3900)$ and $Z_c^{\pm}(4020)$ stand out as two extensively investigated particles, both theoretically and experimentally, with their existence confirmed by multiple collaborations across various decay channels. The $Z_c^{\pm}(3900)$ was the first confirmed charged charmonium-like states, initially observed by the BESIII~\cite{BESIII:2013ris} and Belle~\cite{Belle:2013yex} Collaborations in 2013 in the $J/\psi \pi^{\pm}$ invariant mass distributions of the process $e^+e^- \to \pi^+\pi^-J/\psi$ at $\sqrt{s}=4.26$ GeV. This state was subsequently confirmed by the authors in Ref.~\cite{Xiao:2013iha} in the same process but at $\sqrt{s}=4.17$ GeV using the data collected by the CLEO-c detector. As the heavy quark spin symmetry (HQSS) partner of $Z_c^{\pm}(3900)$, the $Z_c^\pm(4020)$ was observed in the $\pi^\pm h_c$ mass spectrum via $e^+e^- \to \pi^+\pi^- h_c$ process~\cite{BESIII:2013ouc} by the BESIII Collaborations in the same year. In addition to hidden charm process, the BESIII collaboration also observed both states in the $(D^{\ast }\bar{D}^{(\ast)})^\pm$ invariant mass spectrum of the $e^+e^- \to \pi D^{\ast }\bar{D}^{(\ast)}$ processes~\cite{BESIII:2013qmu,BESIII:2013mhi}.

The observed masses of $Z_c(3900)$ and $Z_c(4020)$ lie close to the threshold of $D^*\bar{D}$ and $D^*\bar{D}^*$, respectively. This particular property has led to the proposition of the molecular interpretations with $I=1$. In the framework of the one-boson-exchange model, $Z_c(3900)$ and $Z_c(4020)$ were accommodated as $D^*\bar{D}$ and $D^*\bar{D}^*$ molecular states with $I=1$~\cite{Sun:2012zzd,He:2013nwa}. In Ref.~\cite{Chen:2013omd}, the authors applied QCD sum rule in the $D^*\bar{D}^*$ molecular scenario to reproduce the mass of $Z_c(4020)$. The decay behaviors of $Z_c(3900)/Z_c(4020)$ had also been investigated within the molecular framework in Refs.~\cite{Dong:2013iqa,Wang:2013cya,Li:2013xia,Li:2014pfa,Xiao:2018kfx,Chen:2016byt,Chen:2015igx}. Since these two charmonium-like states had been observed in the hidden-charm final states, the most possible quark components of charged charmonium-like states are $c\bar{c}q\bar{q}$, which indicate that they could be regarded as tetraquark states~\cite{Maiani:2005pe,Nielsen:2006jn,Dubnicka:2010kz,Deng:2014gqa,Deng:2015lca,Brodsky:2014xia,Lebed:2017min}. In addition to QCD exotic states interpretations, $Z_c(3900)$ and $Z_c(4020)$ could also be well reproduced by the initial single pion emission(ISPE) mechanism~\cite{Chen:2013coa,Chen:2013bha,Wang:2013qwa,Chen:2012yr}.

According to SU(3) flavor symmetry, the strange partners of $Z_c$ states with quark content $c\bar{c}s\bar{q}$ should also exist and have in fact been predicted by several theoretical models~\cite{Ebert:2008kb,Ferretti:2020ewe,Lee:2008uy,Dias:2013qga,Chen:2013wca}. In 2021, the BESIII Collaboration observed the strangeness-flavor partner of the $Z_c(3900)$, named $Z_{cs}(3985)$, in the $K^+$ recoil-mass spectrum of the process $e^+e^- \to K^+(D_s^- D^{*0}+D_s^{*-}D^0)$~\cite{BESIII:2020qkh}. Subsequently, a structure, $Z_{cs}(4000)$, with $J^P=1^+$ was observed by the LHCb Collaboration in the $B^+ \to J/\psi \phi K^+$ decay~\cite{LHCb:2021uow}. The resonance parameters of these two states are,
\begin{eqnarray}
	Z_{cs}(3985):&& {\rm M}=3982.5\pm2.1_{-2.6}^{+1.8} ~{\rm MeV},\nonumber \\
	\qquad\qquad\qquad  &&\Gamma=12.8_{-4.4}^{+5.3}\pm3.0 ~{\rm MeV},\nonumber \\
	Z_{cs}(4000):&& {\rm M}=4003.0\pm6.0_{-14.0}^{+4.0} ~{\rm MeV},\nonumber \\
	\qquad\qquad\qquad  &&\Gamma=131.0\pm15\pm26~{\rm MeV}.
\end{eqnarray}
One can find that the masses of these two states are pretty close, while their widths differ significantly. Because of this property, whether $Z_{cs}(3985)$ and $Z_{cs}(4000)$ are the same state remains an open question. Following Refs.~\cite{Yang:2020nrt,Ortega:2021enc,Wu:2021cyc}, we assume them as the same state and use $Z_{cs}$ to refer them in the present work.

The theoretical interpretations of $Z_{cs}$ are in line with those of $Z_c^{\pm}(3900)$ and $Z_c^{\pm}(4020)$, which range from molecular states~\cite{Wang:2020rcx,Ozdem:2021yvo,Meng:2020ihj,Yang:2020nrt,Sun:2020hjw,Wang:2020htx,Dong:2020hxe,Xu:2020evn,Yan:2021tcp}, compact tetraquark states~\cite{Ozdem:2021yvo,Wan:2020oxt,Wang:2020iqt,Jin:2020yjn,Giron:2021sla,Yang:2021zhe}, to kinematic effects~\cite{Wang:2020kej,Ikeno:2020mra}. Given that the mass of $Z_{cs}$ lies near the threshold of $D_s\bar{D}^*/D_s^*\bar{D}$, it is natural to regard it as $D_s^{(*)}\bar{D}^{(*)}$ molecular state~\cite{Wang:2020rcx,Ozdem:2021yvo,Meng:2020ihj,Yang:2020nrt,Sun:2020hjw,Wang:2020htx,Dong:2020hxe,Xu:2020evn,Yan:2021tcp}. However, studies employing the one-boson-exchange(OBE) model do not support $Z_{cs}$ as a hadronic molecular state or a pure molecular state~\cite{Liu:2020nge,Chen:2020yvq}. In the tetraquark scenario, analyses based on QCD sum rules~\cite{Ozdem:2021yvo,Wan:2020oxt,Wang:2020iqt} and quark models\cite{Jin:2020yjn,Giron:2021sla,Yang:2021zhe} suggest that $Z_{cs}$ may instead be a compact tetraquark state. Besides the molecular and tetraquark states, the reflection mechanism~\cite{Wang:2020kej} and threshold effect~\cite{Ikeno:2020mra} were also proposed to reveal the nature of $Z_{cs}$ state.

The inner structure of the $Z_{cs}$ remains elusive. In order to better understand its nature, one might look for more production modes. One of the most important production processes of exotic states is the $B^+$ decay process, where the LHCb Collaboration has observed $Z_{cs}$ in $B^+\to J/\psi \phi K^+$ decays, reporting a fit fraction~\cite{LHCb:2021uow},
\begin{eqnarray}
	\frac{\mathcal{B}[B^+\to Z_{cs}^+\phi \to J/\psi\phi K^+]}{\mathcal{B}[B^+ \to J/\psi \phi K^+]}=(9.4\pm2.1\pm3.4)\%.
\end{eqnarray}
Considering the PDG average of the branching fraction of $B^+ \to J/\psi \phi K^+$ to be $(5.0\pm0.4)\times10^{-5}$~\cite{ParticleDataGroup:2022pth}, one can conclude that the branching fraction of the cascade process is,
\begin{eqnarray}\label{eq:cascade}
 \mathcal{B}[B^+ \to  Z_{cs}^+\phi \to J/\psi K^+ \phi]=(4.6\pm2.0)\times10^{-6}.
\end{eqnarray}
Such a large branching fraction in $B$ decay processes is helpful for us to investigate the property of $Z_{cs}$. In our previous works, we have studied the hidden charm decays~\cite{Wu:2021ezz} and the production of $Z_{cs}$ in $B/B_s$ decays~\cite{Wu:2021cyc}. The branching fraction of the cascade process $B^+ \to  Z_{cs}^+\phi \to J/\psi K^+ \phi$ is estimated to be $(1.86^{+2.12}_{-1.37})\times10^{-6}$ by considering the triangle loop with $D^{(*)}_s \bar{D}^{(*)} D^{(*)}_s$~\cite{Wu:2021cyc}, which is comparable with the experimental measurement from the LHCb Collaboration. According to the Review of Particle Physics (RPP)~\cite{ParticleDataGroup:2022pth}, the branching fractions of $B^+\to \bar{D}^0 D^+_{s1}(2460)$ and $B^+\to \bar{D}^{*0} D^+_{s1}(2460)$ are $(3.1^{+1.0}_{-0.9})\times10^{-3}$ and $(12.0\pm3.0)\times10^{-3}$, respectively, which are in the same order of $B^+\to \bar{D}^{(*0)} D^+_s$.\footnote{In this work, $D^\prime_{s1}(2460)^+$ is treated as $P$-wave charmed-strange meson with $J^{P}=1^+$ and it will be abbreviated as $D^{\prime+}_{s1}$ in the following.} Moreover, the $D^{(*)+}_s D^{(*)+}_s \phi$ vertices have a $P$-wave coupling, while the $D^{\prime+}_{s1}$ couples to the $D^{(*)+}_s \phi$ in S-waves. Thus, the triangle loop with $D^{\prime+}_{s1} \bar{D}^{(*)} D^{(*)}_s$ could also contribute to the process $B^+\to \phi Z^+_{cs}$ and we will revisit the process $B^+\to \phi Z^+_{cs}$ by including the triangle loop with $D^{\prime+}_{s1} \bar{D}^{(*)} D^{(*)}_s$ in this work.

\begin{figure}[t]
	\begin{tabular}{cc}
		\centering
\includegraphics[width=6cm]{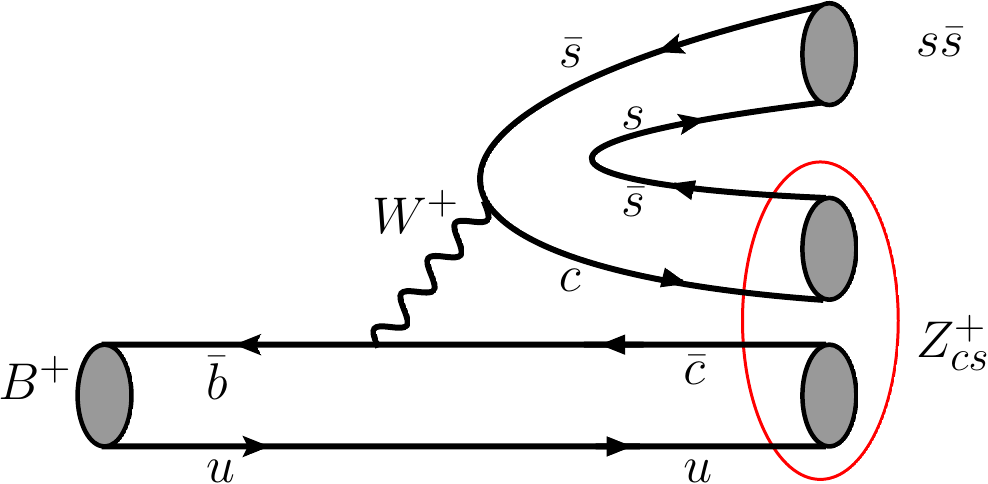}& \\
(a) &\vspace{0.2cm} \\
\includegraphics[width=6cm]{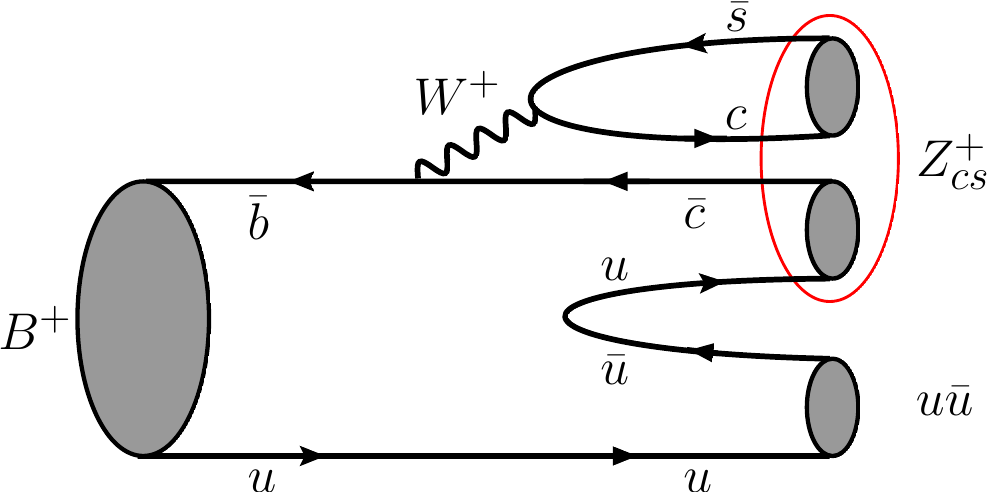}&\\
(b)&\\
	\end{tabular}
	\caption{Diagrams contributing to $B^+ \to Z_{cs}^+ s\bar{s}/u\bar{u}$ at the quark level.}\label{Fig:quarklevel}
\end{figure}

\begin{table}[t]
	\caption{The considered processes in the present work and their branching fractions in PDG average~\cite{ParticleDataGroup:2022pth}\label{Tab:bb}}
	\renewcommand\arraystretch{1.5}
	\begin{tabular}{p{3.5cm}<\centering p{4.5cm}<\centering }
		\toprule[1pt]
		Process & branching fraction  \\
		\midrule[1pt]
		$B^+ \to J/\psi K^+\phi$ & $(5.0\pm0.4)\times10^{-5}$   \\
		$B^+ \to J/\psi K^+\omega$ & $(3.2_{-0.32}^{+0.60})\times10^{-4}$ \\
		$B^+ \to J/\psi K^+\eta$ & $(1.24\pm0.14)\times10^{-4}$  \\
		$B^+ \to J/\psi K^+\eta^\prime$ & $(3.1\pm0.4)\times10^{-5}$ \\
		$B^+ \to J/\psi K^+\pi^0$ & $(1.14\pm0.1)\times10^{-3}$
		  \\
		 $B^+ \to J/\psi K^+\rho^0$ & $-$
		  \\
		\bottomrule[1pt]	
	\end{tabular}
\end{table}

In the present work, we aim to systematically investigate the production of $Z_{cs}$ in $B$ decay with a light unflavored meson. Given that the quark components of $Z_{cs}^+$ are the molecular state composed of $D_s^+ \bar{D}^0$, there are two different production mechanisms at quark level, as shown in Fig.~\ref{Fig:quarklevel}. In Fig.~\ref{Fig:quarklevel}-(a), the anti-bottom quark transits into an anti-charmed quark by emitting a $W^+$ boson, which consequently decays into a charmed quark and an anti-strange quark. The charmed quark hadronize as $D_s^+$ with the anti-strange quark created from the vacuum, while the rest anti-strange quark from $W^+$ decay and strange quark created from vacuum form a light unflavor hadron containing $s\bar{s}$ quark components, such as $\phi$, $\eta$ and $\eta^\prime$. The anticharmed quark coming from the anti-bottom quark and the spectated up quark form a $\bar{D}^0$, then the $\bar{D}^0$ and $D_s^+$ are bounded as the molecular state $Z_{cs}^+$. A similar mechanism can be found in Fig.~\ref{Fig:quarklevel}-(b). Differ from the mechanism in Fig~~\ref{Fig:quarklevel}-(a), the quark pair created from the vacuum is $u\bar{u}$, and the $u\bar{u}$ in the final states could   hadronize as $\pi^0$, $\eta$, $\eta^\prime$, $\omega$, and $\rho^0$ mesons.

In the present work, we evaluate the possibility of searching $Z_{cs}^+$ in the $J/\psi K^+$  invariant mass distributions of the decay processes $B^+ \to J/\psi K^+ \mathbb{M}$, where $\mathbb{M}$ can be $\phi/\omega/\rho^0/\pi^0/\eta/\eta^\prime$. Fortunately, all the above the processes except for $B^+ \to J/\psi K^+ \rho^0$ have been measured experimentally, and the branching fractions range from $10^{-5}$ and $10^{-3}$, which are listed in Table ~\ref{Tab:bb}. Using the branching fraction of $Z^+_{cs}\to J/\psi K^+$ estimated in Ref.~\cite{Wu:2021ezz}, we will revisit the fit fraction of $Z_{cs}$ in $B^+ \to J/\psi K^+\phi$ and predict the fit fractions of $Z_{cs}$ in $B^+ \to J/\psi K^+/\omega/\eta^{(\prime)}/\pi^0$. The branching fraction of $B^+ \to J/\psi K^+ \rho^0$ remains unclear up till now, but we still consider it in the present work.

The rest of this work is organized as follows. After the introduction, we present the model used in the estimation of $Z_{cs}$ productions. The numerical results and the relevant discussions  are given in Section.~\ref{Sec:Num}, and Section.~\ref{Sec:sum} is devoted to a brief summary.

\section{Theoretical framework}
\label{Sec:Method}

\begin{table}[t]
	\caption{All the loops considered in the present work contribute to $B^+ \to Z_{cs}^+ \phi/\omega/\rho^0/\eta/\eta^\prime/\pi^0.$ \label{Tab:loops}}
	\renewcommand\arraystretch{1.5}
	\begin{tabular}{p{2.6cm}<\centering p{5.7cm}<\centering }
		\toprule[1pt]
		Process & Loops  \\
		\midrule[1pt]
		$B^+ \to Z_{cs}^+ \phi$ & $D_{s}^{+} \bar{D}^{0} D_{s}^{*+},~D_{s}^{+} \bar{D}^{*0} D_{s}^{+},~D_{s}^{*+} \bar{D}^{0} D_{s}^{*+}$\\ & $D_{s}^{*+} \bar{D}^{*0} D_{s}^{+},~ D_{s1}^{\prime +} \bar{D}^{0} D_{s}^{*+},~D_{s1}^{\prime +} \bar{D}^{*0} D_{s}^{+},$ \\
		\midrule[1pt]
		$B^+ \to Z_{cs}^+ \omega/\rho^0$ & $\bar{D}^{0} D_{s}^{+} \bar{D}^{*0},~\bar{D}^{*0} D_{s}^{+} \bar{D}^{*0},~$ \\ & $\bar{D}^{0} D_{s}^{*+} \bar{D}^{0},~\bar{D}^{*0} D_{s}^{*+} \bar{D}^{0},$ \\
		\midrule[1pt]
		$B^+ \to Z_{cs}^+ \eta/\eta^\prime$ & $D_{s}^{+} \bar{D}^{0} D_{s}^{*+},~D_{s}^{*+} \bar{D}^{0} D_{s}^{*+},~D_{s}^{*+} \bar{D}^{*0} D_{s}^{+},~D_{s1}^{\prime +} \bar{D}^{0} D_{s}^{*+}$ \\ & $\bar{D}^{0} D_{s}^{+} \bar{D}^{*0},~\bar{D}^{*0} D_{s}^{+} \bar{D}^{*0},~\bar{D}^{*0} D_{s}^{*+} \bar{D}^{0}$\\
		\midrule[1pt]
		$B^+ \to Z_{cs}^+ \pi^0$ & $\bar{D}^{0} D_{s}^{+} \bar{D}^{*0},~\bar{D}^{*0} D_{s}^{+} \bar{D}^{*0},~\bar{D}^{*0} D_{s}^{*+} \bar{D}^{0}$  \\
		\bottomrule[1 pt]	
	\end{tabular}
\end{table}

\begin{figure}[t]
	\begin{tabular}{cccc}
		\centering
		\includegraphics[width=4cm]{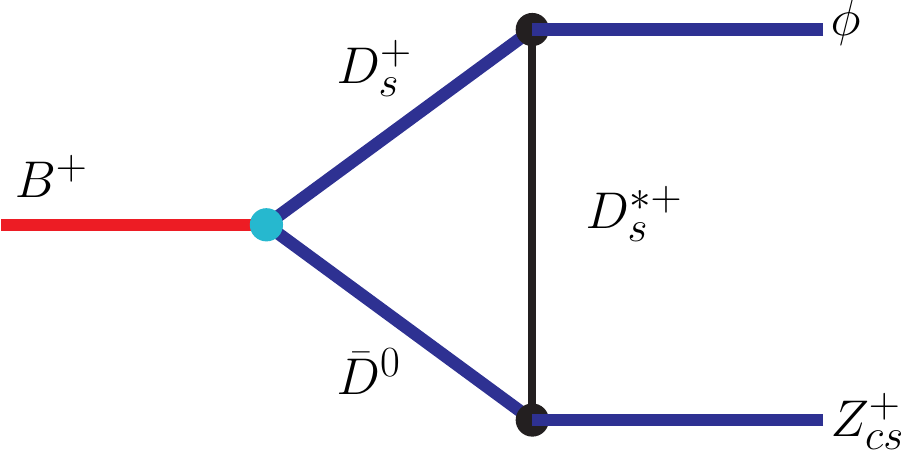}&
		\includegraphics[width=4cm]{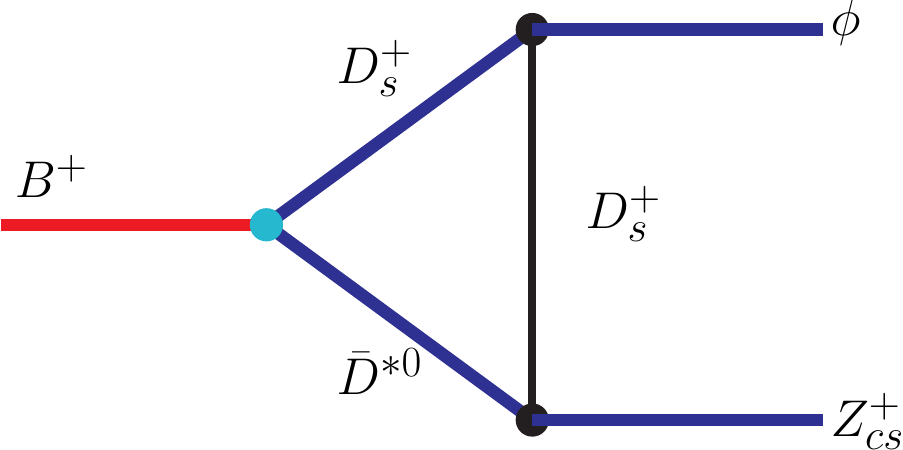}\\
		$(a)$ & $(b)$ \\ \\
		\includegraphics[width=4cm]{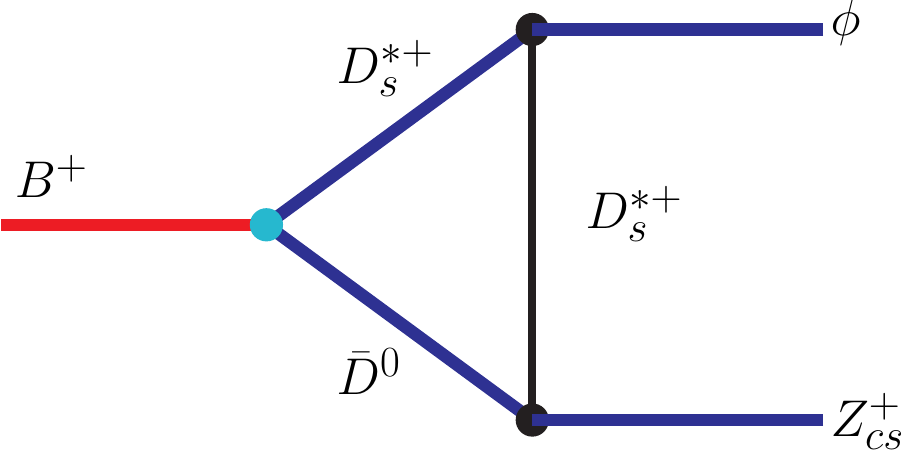}&
		\includegraphics[width=4cm]{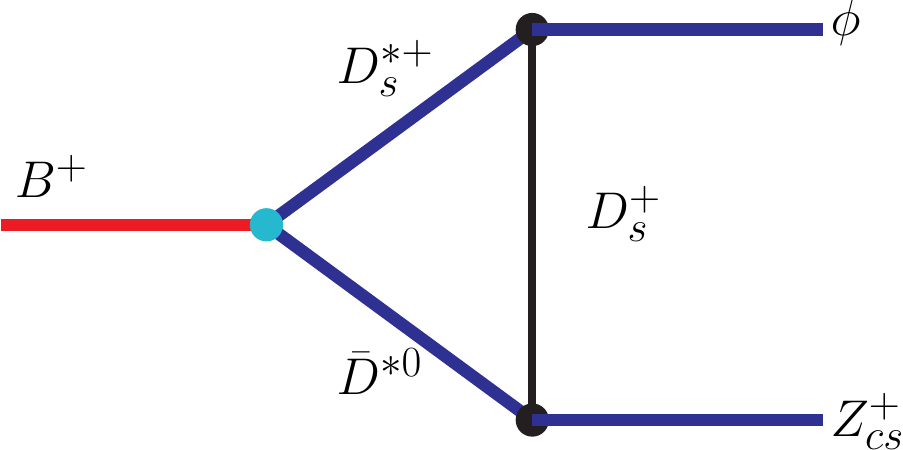}\\
		$(c)$ & $(d)$ \\ \\
		\includegraphics[width=4cm]{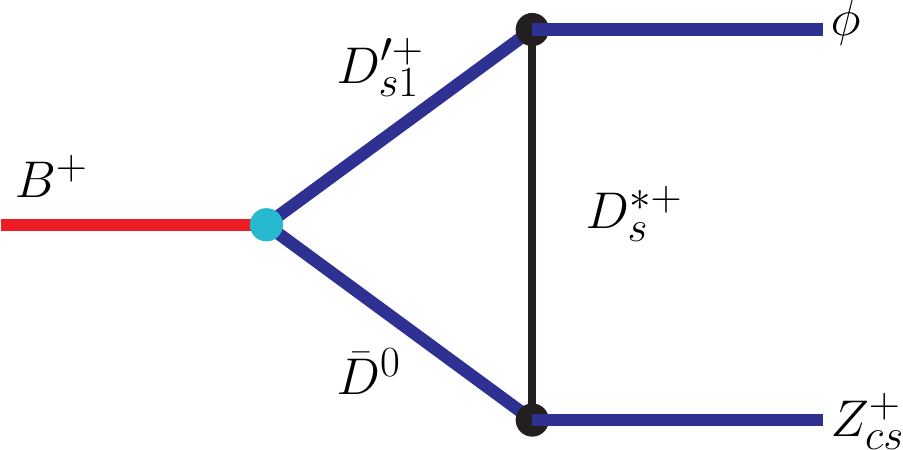}&
		\includegraphics[width=4cm]{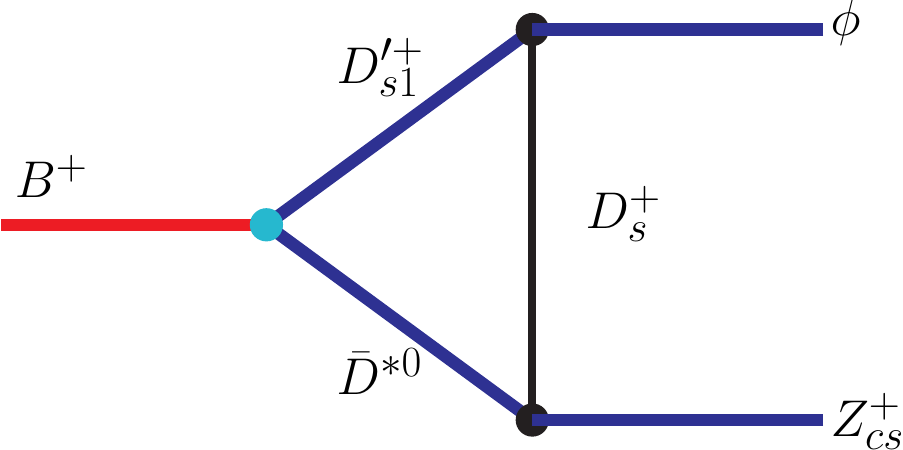}&\\
		\\
		$(e)$  & $(f)$ \\
	\end{tabular}
	\caption{Diagrams contributing to $B^+ \to Z_{cs}^+ \phi$ at the hadron level.}\label{Fig:tri-phi}
\end{figure}

In the present work, we estimate the production processes at the hadron level, where the molecular state $Z_{cs}^+$ could be produced in $B^+$ meson decay through the triangle loop mechanism. In our model, the $B^+$ meson first decays into a charm-strange meson and an anti-charmed meson through weak interaction, and they transit into $Z_{cs}$ and a light unflavor meson by exchanging a proper charm-strange or charmed meson. In Table~\ref{Tab:loops}, we collected the possible meson loops contributing to the discussed processes, where $\mathbb{M}_1 \mathbb{M}_2 \mathbb{M}_3$ refer to the meson loop constructed by $\mathbb{M}_1$, $ \mathbb{M}_2$ and $\mathbb{M}_3$. Here, we take $B^+ \to  Z_{cs}^+ \phi$ as an example, the relevant diagrams are presented in Fig.~\ref{Fig:tri-phi}. We employe the effective Lagrangian approach to calculate the triangle loop diagrams. In the following subsections, we present the relevant effective Lagrangian and decay amplitude for calculating these production processes.

\subsection{Effective Lagrangian}
In the present work, $Z_{cs}^+$ is assumed to be a $S$-wave bound state of $D_s^+\bar{D}^{*0}+D_s^{*+}\bar{D}^0$ with $I(J^P)=\frac{1}{2}(1^+)$, which is,
\begin{eqnarray}
	Z_{cs}^+=\frac{1}{\sqrt{2}}\left(|D_s^{*+}\bar{D}^0\rangle+|D_s^+\bar{D}^{*0}\rangle\right),
\end{eqnarray}
and the effective Lagrangian depicting the interaction of $Z_{cs}$ with its components can be,
\begin{eqnarray}
	\mathcal{L}_{Z_{cs}}=\frac{g_{Z_{cs}}}{\sqrt{2}}Z_{cs}^\mu\left(D_{s\mu}^*\bar{D} + D_s\bar{D}_\mu^*\right),
\end{eqnarray}
where $g_{Z_{cs}}$ is the effective coupling constant.

As for the $B^+$ weak decay vertex,  we utilized the parametrized hadronic matrix elements obtained by the effective Hamiltonian at the quark level, which are~\cite{Cheng:2003sm,Soni:2021fky},
\begin{eqnarray}
\label{eq:weak transition}
&&\left\langle0|J_\mu|P(p_1)\right\rangle = -if_p p_{1\mu}\;,\nonumber\\
&&\left\langle0|J_\mu|V(p_1,\epsilon)\right\rangle =f_V \epsilon_\mu m_V \;,\nonumber\\
&&\left\langle P(p_2)|J_\mu|B(p)\right\rangle \nonumber\\&&\quad=\left[P_\mu-\frac{m^2_{B}-m^2_P}{q^2}q_\mu\right]F_1\left(q^2\right)
+\frac{m^2_{B}-m^2_P}{q^2}q_\mu F_0\left(q^2\right) \;,\nonumber\\
&&\left\langle V(p_2,\epsilon)|J_\mu|B(p)\right\rangle \nonumber\\
&&\quad=\frac{i\epsilon^\nu}{m_{B}+m_V}\Bigg\{i\varepsilon_{\mu\nu\alpha\beta}P^\alpha q^\beta V\left(q^2\right)\nonumber\\
&&\quad -(m_{B}+m_V)^2 g_{\mu\nu}A_1\left(q^2\right) +P_\mu P_\nu A_2\left(q^2\right)\nonumber\\
&&\quad+2m_V(m_{B}+m_V)\frac{P_\nu q_\mu}{q^2}\left[A_3\left(q^2\right)-A_0\left(q^2\right)\right] \Bigg\}\;,
\end{eqnarray}
with $J_{\mu}=\bar{q}_1 \gamma_{\mu}(1-\gamma5)q_2$, $P_{\mu}=p_\mu+p_{2\mu}$ and $q_{\mu}=p_{\mu}-p_{2\mu}$.  $V,A_{0,1,2,+,-}\left(q^2\right)$ and $F_{0,1,+,-}\left(q^2\right)$ are the weak transition form factors, while $A_3\left(q^2\right)$ is the linear combination of form factors $A_1\left(q^2\right)$ and $A_2\left(q^2\right)$, which is~\cite{Cheng:2003sm},
\begin{eqnarray}
A_3\left(q^2\right)=\frac{m_B+m_{V}}{2m_{V}}A_1\left(q^2\right)-\frac{m_B-m_{V}}{2m_{V}}A_2\left(q^2\right).
\end{eqnarray}
More details of these form factors will be discussed in the following section.

The amplitudes of the $B^+$ weak transition could be constructed by the products of two hadronic matrix elements. Here, we take the vertex of $B^+ \rightarrow D_{s1}^+\bar{D}^*$ as an example, the amplitude reads,
\begin{eqnarray}
&&\mathcal{A}(B^+ \to D_{s1}^+ \bar{D}^*) =	\mathcal{A}_{\mu\nu }^{B^+ \to D_{s1}^+ \bar{D}^*} \epsilon_{D_{s1}}^\mu \epsilon_{\bar{D}^\ast}^\nu \nonumber\\
&&\qquad = \frac{G_F}{\sqrt{2}}V_{cb}^*V_{cs}a_1	\left \langle D_{s1}^+ \left | J_\mu^\dagger \right |0  \right \rangle 	\left \langle \bar{D}^* \left | J_\mu \right |B^+  \right \rangle \epsilon_{D_{s1}}^\mu \epsilon_{\bar{D}^\ast}^\nu, \hspace{0.5cm}
\end{eqnarray}
where $G_F$ is the Fermi constant, $V_{cb}^*$ and $V_{cs}$ are the CKM matrix elements. $a_1=c_1^{eff}+c_2^{eff}/N_c$ with $c_{1,2}^{eff}$ to be effective Wilson coefficients obtained by the factorization approach\cite{Bauer:1986bm}. In the present estimations, we adopt $G_F=1.166 \times 10^{-5} {\rm GeV}^{-2}$, $V_{cb}=0.041$, $V_{cs}=0.987$ and $a_1=1.05$ as in Refs.~\cite{ParticleDataGroup:2020ssz,Ali:1998eb,Ivanov:2006ni}.

The interaction of light unflavored mesons and heavy-light mesons can be described by effective Lagrangians constructed based on heavy quark limit and chiral symmetry. The relevant effective Lagrangians are given by,
\begin{eqnarray}
	&&\mathcal{L}_{\mathcal{D}^*\mathcal{D}^{(*)}\mathcal{P}}=-i g_{\mathcal{D}^{*} \mathcal{D}\mathcal{P}}\left(\mathcal{D}_{i}^{\dagger} \partial_{\mu} \mathcal{P}_{i j} \mathcal{D}_{j}^{* \mu}-\mathcal{D}_{i}^{* \mu \dagger} \partial_{\mu} \mathcal{P}_{i j} \mathcal{D}_{j}\right) \nonumber\\
	&&\qquad +\frac{1}{2} g_{\mathcal{D}^{*} \mathcal{D}^{*} \mathcal{P}} \varepsilon_{\mu \nu \alpha \beta}\mathcal{D}_{i}^{* \mu \dagger} \partial^{\nu} \mathcal{P}_{i j} \stackrel{\leftrightarrow}{\partial^{\alpha} }\mathcal{D}_{j}^{* \beta}+\text { H.c. },\nonumber\\
	&&\mathcal{L}_{\mathcal{D}^{(*)}\mathcal{D}^{(*)}\mathcal{V}}=-i g_{\mathcal{D} \mathcal{D} \mathcal{V}} \mathcal{D}_{i}^{\dagger}\stackrel{\leftrightarrow}{\partial_{\mu}} \mathcal{D}^{j}\left(\mathcal{V}^{\mu}\right)_{j}^{i} \nonumber\\		&&\qquad -2 f_{\mathcal{D}^{*} \mathcal{D} \mathcal{V}} \varepsilon_{\mu \nu \alpha\beta}(\partial^{\mu}\mathcal{V}^{\nu})_{j}^{i}(\mathcal{D}_{i}^{\dagger} \stackrel{\leftrightarrow}{\partial^{\alpha}} \mathcal{D}^{* \beta j}-\mathcal{D}_{i}^{* \beta \dagger} \stackrel{\leftrightarrow}{\partial^{\alpha}} \mathcal{D}^{j})\nonumber\\
	&&\qquad +i g_{\mathcal{D}^{*} \mathcal{D}^{*} \mathcal{V}} \mathcal{D}_{i}^{* \nu \dagger} \stackrel{\leftrightarrow}{\partial_{\mu}} \mathcal{D}_{v}^{* j}\left(\mathcal{V}^{\mu}\right)_{j}^{i} \nonumber\\
	&&\qquad +4 i f_{\mathcal{D}^{*} \mathcal{D}^{*} \mathcal{V}} \mathcal{D}_{i \mu}^{* \dagger}\left(\partial^{\mu} \mathcal{V}^{\nu}-\partial^{\nu} \mathcal{V}^{\mu}\right)_{j}^{i} \mathcal{D}_{\nu}^{* j}+\text { H.c. },\nonumber\\
	&&\mathcal{L}_{\mathcal{D}_1^{\prime}\mathcal{D}^{*}\mathcal{P}}=g_{\mathcal{D}_1^{\prime}\mathcal{D}^{*}\mathcal{P}}\Big[\mathcal{D}_{1b\mu}^\prime\stackrel{\leftrightarrow}{\partial^{\mu}}\mathcal{D}_{a\nu}^{*\dagger}\partial^\nu\mathcal{P}_{ba}\nonumber\\
	&&\qquad -\mathcal{D}_{1b\mu}^\prime\stackrel{\leftrightarrow}{\partial^{\nu}}\mathcal{D}_{a}^{*\mu\dagger}\partial_\nu\mathcal{P}_{ba}+\mathcal{D}_{1b\mu}^\prime \stackrel{\leftrightarrow}{\partial^{\nu}} \mathcal{D}_{\nu}^{*\dagger} \mathcal{P}_{ba} \Big], \nonumber\\
	&&\mathcal{L}_{\mathcal{D}_1^{\prime}\mathcal{D}^{(*)}\mathcal{V}}=g_{\mathcal{D}_1^{\prime}\mathcal{D}\mathcal{V}} \mathcal{D}_b \mathcal{D}_{1\mu a}^{\prime\dagger} \left(\mathcal{V}^\mu\right)_{ba} ,  \nonumber\\
	&&\qquad + i g_{\mathcal{D}_1^{\prime}\mathcal{D}^{*}\mathcal{V}} \varepsilon_{\mu\nu\alpha\beta}\mathcal{D}^{*\mu}\stackrel{\leftrightarrow}{\partial^{\beta}}\mathcal{D}_1^{\prime\alpha\dagger} \left(\mathcal{V}^\nu\right)_{ba}	,
\end{eqnarray}
where $\mathcal{D}_1^{\prime}=\left( D_1^{\prime}(2430)^0,D_1^{\prime}(2430)^+,D_{s1}^{\prime}(2460)^+ \right)$, $\mathcal{D}^{(*)\dagger}=\left(D^{(*)0}, D^{(*)+}, D_s^{(*)+}\right)$  and $A\overleftrightarrow{\partial_{\mu}}B = A \partial^\mu B - B \partial^\mu A$. $\mathcal{P}$ and $\mathcal{V}$ are $3\times3$ matrices representing pseudoscalar and vector mesons, with their concrete forms being
\begin{eqnarray}\label{eq:pscalar matrix}
	\mathcal{P} &=&
	\left(\begin{array}{ccc}
		\frac{\pi^{0}}{\sqrt 2}+\alpha\eta+\beta\eta^\prime &\pi^{+} &K^{+}\\
		\pi^{-} &-\frac{\pi^{0}}{\sqrt2}+\alpha\eta+\beta\eta^\prime&K^{0}\\
		K^{-} &\bar K^{0} &\gamma\eta+\delta\eta^\prime
	\end{array}\right),
\end{eqnarray}
\begin{eqnarray}\label{eq:matrix V}
	\mathcal{V} &=& \left(\begin{array}{ccc}\frac{\rho^0} {\sqrt {2}}+\frac {\omega} {\sqrt {2}}&\rho^+ & K^{*+} \\
		\rho^- & -\frac {\rho^0} {\sqrt {2}} + \frac {\omega} {\sqrt {2}} & K^{*0} \\
		K^{*-}& {\bar K}^{*0} & \phi \\
	\end{array}\right),
\end{eqnarray}
where the parameters $\alpha$, $\beta$, $\gamma$ and $\delta$ in Eq.~\eqref{eq:pscalar matrix} related to the mixing angle $\theta$ are defined as
\begin{eqnarray}
	\alpha&=&\frac{\cos\theta-\sqrt{2}\sin\theta}{\sqrt{6}},\ \quad \beta=\frac{\sin\theta+\sqrt{2}\cos\theta}{\sqrt{6}},\nonumber\\
	\gamma&=&\frac{-2\cos\theta-\sqrt{2}\sin\theta}{\sqrt{6}},\ \delta=\frac{-2\sin\theta+\sqrt{2}\cos\theta}{\sqrt{6}},
\end{eqnarray}
with the mixing angle $\theta=-14.1^\circ$ \cite{MARK-III:1988crp,DM2:1988bfq}.

\subsection{Decay Amplitude}

With the effective Lagrangians discussed in the previous section, we can now obtain the amplitudes corresponding to the processes shown in Fig.~\ref{Fig:tri-phi}, which read, 
\begin{eqnarray}
 &&\mathcal{M}_{a}= i^3 \int\frac{d^4q}{(2\pi)^4}\Big[\mathcal{A}^{B^+\rightarrow D_{s}^+ \bar{D}^0}(p_1,p_2)\Big]\Big[2 f_{D^*DV}\varepsilon_{\mu_2\delta\alpha_2\beta}\nonumber\\
 &&\qquad (i p_3^{\mu_2})(p_1+q)^{\alpha_2}\Big]\Big[i g_{Z_{cs}D_s^*D} g_{\delta\phi}\Big] \Big[\frac{1}{p_1^2-m_1^2}\Big]\Big[\frac{1}{p_2^2-m_2^2}\Big] \nonumber\\
 &&\qquad \Big[\frac{-g^{\beta\phi}+q^{\beta} q^{\phi}/m_q^2}{q^2-m_q^2}\Big]\mathcal{F}^2(q^2,m_q^2), \nonumber\\
&&\mathcal{M}_{b}= i^3 \int\frac{d^4q}{(2\pi)^4}\Big[\mathcal{A}_{\nu}^{B^+\rightarrow D_{s}^+ \bar{D}^{*0}}(p_1,p_2)\Big]\Big[-g_{DDV}(p_1+q)_\delta \Big]\nonumber\\
 &&\qquad \Big[i g_{Z_{cs}D_s^*D} g_{\delta\rho}\Big] \Big[\frac{1}{p_1^2-m_1^2}\Big]\Big[\frac{-g^{\nu\rho}+p_2^{\nu} p_2^{\rho}/m_2^2}{p_2^2-m_2^2}\Big] \nonumber\\
 &&\qquad \Big[\frac{1}{q^2-m_q^2}\Big]\mathcal{F}^2(q^2,m_q^2), \nonumber\\
&&\mathcal{M}_{c}= i^3 \int\frac{d^4q}{(2\pi)^4}\Big[\mathcal{A}_{\mu}^{B^+\rightarrow D_{s}^{*+} \bar{D}^0}(p_1,p_2)\Big]\Big[i g_{D^*D^*V}(p_1+q)_\delta g_{\alpha\beta}\nonumber\\
 &&\qquad+4i f_{D^*D^*V}((i p_{3\beta})g_{\alpha\delta}-(i p_{3\alpha})g_{\delta\beta}) \Big]\Big[i g_{Z_{cs}D_s^*D} g_{\delta\phi}\Big]\nonumber\\
 &&\qquad \Big[\frac{-g^{\mu\alpha}+p_1^{\mu} p_1^{\alpha}/m_1^2}{p_1^2-m_1^2}\Big]\Big[\frac{1}{p_2^2-m_2^2}\Big]\Big[\frac{-g^{\beta\phi}+q^{\beta} q^{\phi}/m_q^2}{q^2-m_q^2}\Big]\nonumber\\
 &&\qquad \mathcal{F}^2(q^2,m_q^2), \nonumber\\
  &&\mathcal{M}_{d}= i^3 \int\frac{d^4q}{(2\pi)^4}\Big[\mathcal{A}_{\mu\nu}^{B^+\rightarrow D_{s}^{*+} \bar{D}^{*0}}(p_1,p_2)\Big]\Big[-2 f_{D^*DV}\varepsilon_{\mu_2\delta\alpha_2\alpha}\nonumber\\
 &&\qquad (i p_3^{\mu_2})(p_1+q)^{\alpha_2} \Big]\Big[i g_{Z_{cs}D_s^*D} g_{\delta\rho}\Big] \Big[\frac{-g^{\nu\alpha}+p_1^{\mu} p_1^{\alpha}/m_1^2}{p_1^2-m_1^2}\Big]\nonumber\\
 &&\qquad\Big[\frac{-g^{\nu\rho}+p_2^{\nu} p_2^{\rho}/m_2^2}{p_2^2-m_2^2}\Big] \Big[\frac{1}{q^2-m_q^2}\Big]\mathcal{F}^2(q^2,m_q^2), \nonumber\\
 &&\mathcal{M}_{e}= i^3 \int\frac{d^4q}{(2\pi)^4}\Big[\mathcal{A}_{\mu}^{B^+\rightarrow D_{s1}^{\prime+} \bar{D}^0}(p_1,p_2)\Big]\Big[i g_{D_1^\prime D^* V}\varepsilon_{\beta\delta\alpha\beta_2}\nonumber\\
 &&\qquad (p_1+q)^{\beta_2} \Big] \Big[i g_{Z_{cs}D_s^*D} g_{\delta\phi}\Big] \Big[\frac{-g^{\mu\alpha}+p_1^{\mu} p_1^{\alpha}/m_1^2}{p_1^2-m_1^2}\Big]\Big[\frac{1}{p_2^2-m_2^2}\Big] \nonumber\\
 &&\qquad \Big[\frac{-g^{\beta\phi}+q^{\beta} q^{\phi}/m_q^2}{q^2-m_q^2}\Big]\mathcal{F}^2(q^2,m_q^2), \nonumber\\
 &&\mathcal{M}_{f}= i^3 \int\frac{d^4q}{(2\pi)^4}\Big[\mathcal{A}_{\mu\nu}^{B^+\rightarrow D_{s1}^{\prime+} \bar{D}^{*0}}(p_1,p_2)\Big]\Big[i g_{D_1^\prime D V} g_{\delta\alpha}\Big]\nonumber\\
 &&\qquad\Big[i g_{Z_{cs}D_s^*D} g_{\delta\rho}\Big] \Big[\frac{-g^{\nu\alpha}+p_1^{\mu} p_1^{\alpha}/m_1^2}{p_1^2-m_1^2}\Big]\nonumber\\
 &&\qquad\Big[\frac{-g^{\nu\rho}+p_2^{\nu} p_2^{\rho}/m_2^2}{p_2^2-m_2^2}\Big] \Big[\frac{1}{q^2-m_q^2}\Big]\mathcal{F}^2(q^2,m_q^2).  \label{Eq:amp}
\end{eqnarray}
In the above amplitudes, a form factor $\mathcal{F}(q^2,m^2)$ in the momopole form is introduced to compensate for the off-shell effect and avoid ultraviolet divergence in loop integrals. Its concrete form is,
\begin{equation}
	\mathcal{F}\left(q^2,m^2\right) = \frac{m^2-\Lambda^2}{q^2-\Lambda^2}, \label{Eq:FF}
\end{equation}
where the parameter $\Lambda$ can be reparameterized as ${ \Lambda} = m + \alpha {\rm \Lambda_{QCD}}$, where $ \Lambda_{\rm QCD}$ = 220 MeV. $\alpha$ is the model parameter~\cite{Cheng:2004ru}, which should be of order unity. 

Then, the amplitude of $B^+ \to z_{cs}^+ \phi$ reads,
\begin{eqnarray}
\mathcal{M}_{B^+\to Z_{cs}^+ \phi} =\mathcal{M}_a +\mathcal{M}_b+\mathcal{M}_c+\mathcal{M}_d+\mathcal{M}_e+\mathcal{M}_f,
\end{eqnarray}
and the partial width of $B^+ \to z_{cs}^+ \phi$ can be estimated by,
\begin{eqnarray}
	&&\Gamma_{B^+ \to Z_{cs}^+ \phi} = \frac{1}{8\pi} \frac{|\vec{p}\,|}{m_{B^+}^2}\overline{\left|\mathcal{M}_{B^+\to Z_{cs}^+ \phi}\right|^2} ,
\end{eqnarray}
where $\vec{p}$ is the momentum of final states in the rest frame of
$B^+$, and the overline above the amplitude indicates the sum over the spins of final states.

\begin{table}
  \caption{\label{Tab:F0ab}Values of parameters $F(0)$, $a$ and $b$ in form factors~\cite{Cheng:2003sm}.}
\renewcommand\arraystretch{1.5}
	\begin{tabular}{p{1.9cm}<\centering p{1.9cm}<\centering p{1.9cm}<\centering p{1.9cm}<\centering}
		\toprule[1pt]
		Parameter &$F(0)$ &$a$ &$b$ \\
		\midrule[1pt]
  $F_0$ &0.67 &0.65 &0.00 \\
  $F_1$ &0.67 &1.25 &0.39 \\
		$A_0$ &0.64 &1.30 &0.31 \\
		$A_1$ &0.63 &0.65 &0.02 \\
		$A_2$ &0.61 &1.14 &0.52 \\
		$V$ &0.75 &1.29 &0.45 \\
		\bottomrule[1pt]	
	\end{tabular}
\end{table}

\begin{table}
\caption{\label{Tab:L1L2}Values of parameters $\Lambda_1$ and $\Lambda_2$ obtained by fitting Eq.~\eqref{eq:FQlambda} with Eq.~\eqref{eq:FQzeta}.}
\renewcommand\arraystretch{1.5}
	\begin{tabular}{p{1.0cm}<\centering p{1.0cm}<\centering p{1.0cm}<\centering p{1.0cm}<\centering p{1.0cm}<\centering p{1.0cm}<\centering p{1.0cm}<\centering}
		\toprule[1pt]
		Parameter &$F_{0}$ &$F_{1}$ &$A_{0}$ &$A_{1}$ &$A_{2}$ &$V$ \\
		\midrule[0.8pt]
		$\Lambda_1$ &7.75 &6.53 &6.5 &8.95 &6.65 &6.30 \\
	    $\Lambda_2$ &11.00 &6.84 &7.1 &8.75 &7.55 &7.10 \\
        \bottomrule[1pt]
	\end{tabular}
\end{table}

\section{Numerical Results and Discussions}
\label{Sec:Num}
\subsection{Coupling Constants}
Before presenting the numerical results, some relevant constants should be clarified. The decay constants in Eq.~\eqref{eq:weak transition} are taken as $f_{D_s}=250$ MeV, $f_{D_s^*}=272$ MeV and $f_{D_{s1}^\prime}=158$ MeV from Refs.~\cite{FlavourLatticeAveragingGroup:2019iem,Hwang:2004kga}. In addition, the form factors in Eq.~\eqref{eq:weak transition} are usually estimated in the quark model and are known only in the spacelike region. Some methods like analytic continuation could cover the timelike region where the physical decay processes occur. In Refs.~\cite{Cheng:2003sm,Soni:2021fky}, the form factors are parameterized in the form,
\begin{equation}\label{eq:FQzeta}
	F\left(q^2\right) = \frac{F(0)}{1-a\zeta+b\zeta^2},
\end{equation}
with $\zeta = Q^2/m_{B}^2$. The relevant parameters $F(0)$, $a$, and $b$ for each form factor are collected in Table~\ref{Tab:F0ab}. To avoid ultraviolet divergence in the loop integrals and to evaluate the loop integrals via Feynman parameterization method, we further parameterize these form factors in the form,
\begin{equation}\label{eq:FQlambda}
	F\left(q^2\right) = F(0)\frac{\Lambda_1^2}{Q^2-\Lambda_1^2}\frac{\Lambda_2^2}{Q^2-\Lambda_2^2}.
\end{equation}
By fitting Eq.~\eqref{eq:FQlambda} with Eq.~\eqref{eq:FQzeta}, we can obtain the values of $\Lambda_1$ and $\Lambda_2$ for each form factor, which are listed in Table~\ref{Tab:L1L2}.

Considering the chiral symmetry and heavy quark limit, the coupling constants of charmed meson with pseudoscalar and vector meson can be related to gauge couplings by~\cite{Isola:2003fh,Falk:1992cx, Casalbuoni:1996pg},
\begin{eqnarray}
	&&g_{D^*DP} = \frac{2g}{f_{\pi}}\sqrt{m_{D}m_{D^*}},~~~g_{D^*D^*P} = \frac{g_{D^*DP}}{\sqrt{m_{D^*}m_{D^*}}}, \nonumber\\
	&&g_{DDV} = g_{D^*D^*V} = \frac{\beta g_V}{\sqrt{2}},~~~f_{D^*DV} = \frac{ f_{D^*D^*V}}{m_{D^*}}=\frac{\lambda g_V}{\sqrt{2}},\nonumber\\
	&&g_{D_1^\prime D^*P} = \frac{2h}{f_{\pi}},\nonumber\\
	&&g_{D_1^\prime D V}=-\sqrt{2}g_V \zeta \sqrt{m_D m_{D_1^\prime}},~~ g_{D_1^\prime D^* V}=\sqrt{2}g_V \zeta,
\end{eqnarray}
where the parameter $\beta=0.9$, $\zeta=0.1$, $g_V=m_\rho / f_\pi$ with $f_\pi=0.132$ GeV and $h=-0.56$ \cite{Casalbuoni:1996pg}. By matching the
form factor obtained from the light-cone sum rules and lattice QCD, one can obtain the parameter $\lambda = 0.56~{\rm GeV^{-1}}$ and $g = 0.59$\cite{Isola:2003fh}.

In addition to the above coupling constants, the coupling constant $g_{Z_{cs}D_s^\ast D}$ was investigated the decay properties of $Z_{cs}$~\cite{Wu:2021ezz},  and the coupling constant $g_{Z_{cs}D_s^*D}$ was determined to be $6.0 \sim 6.7$, which is weakly dependent on the model parameter. In the following, we take $g_{Z_{cs}D_s^*D}=6.0$ as in Ref.~\cite{Wu:2021cyc}, which investigated the production of $Z_{cs}$ in $B$ and $B_s$ decays.

\begin{table}
	\caption{The estimated branching fractions for all processes (in order of $10^{-4}$). \label{Tab:br}}
	\renewcommand\arraystretch{1.5}
	\begin{tabular}{p{3.5cm}<\centering p{4.5cm}<\centering }
		\toprule[1pt]
		Process & branching fraction \\
		\midrule[1pt]
		$B^+ \to Z_{cs}^+\phi$ & $1.75_{-0.50}^{+0.46}$ \\
		$B^+ \to Z_{cs}^+\omega$ & $1.03_{-0.34}^{+0.36}$ \\
		$B^+ \to Z_{cs}^+\rho^0$ & $1.06_{-0.36}^{+0.36}$   \\
		$B^+ \to Z_{cs}^+\eta$ & $0.29_{-0.12}^{+0.14}$  \\
		$B^+ \to Z_{cs}^+\eta^\prime$ & $0.53_{-0.22}^{+0.27}$  \\
		$B^+ \to Z_{cs}^+\pi^0$ & $0.30_{-0.12}^{+0.13}$ \\
		\bottomrule[1pt]	
	\end{tabular}
\end{table}

\begin{figure*}[t]
	\centering
\includegraphics[width=18 cm]{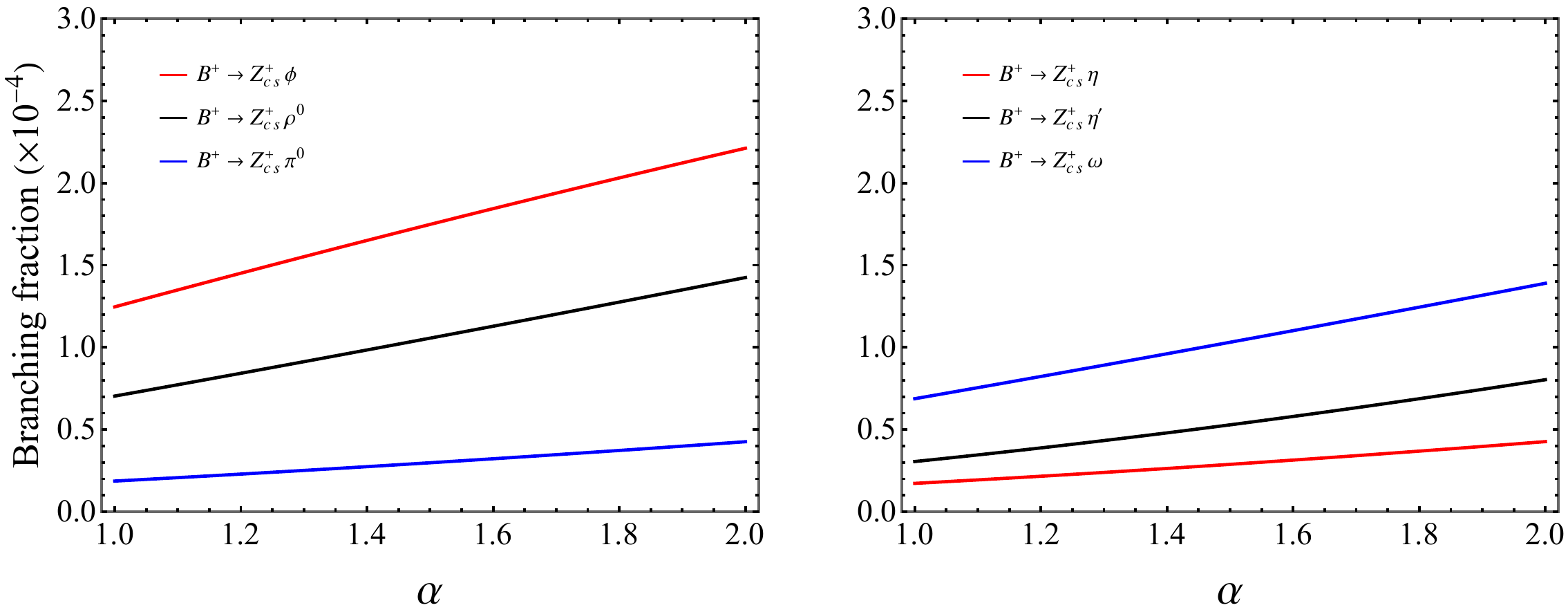}
	\caption{The branching fractions fo $B^+ \to Z_{cs}^+ \mathbb{M}$ depending on the model parameter $\alpha$, where $\mathbb{M}$ can be $\phi/\omega/\rho^0/\pi^0/\eta/\eta^\prime$.}
	\label{Fig:br}
\end{figure*}

\subsection{Branching Fractions}

\begin{figure}[tb]
	\centering
	\includegraphics[width=8.5cm]{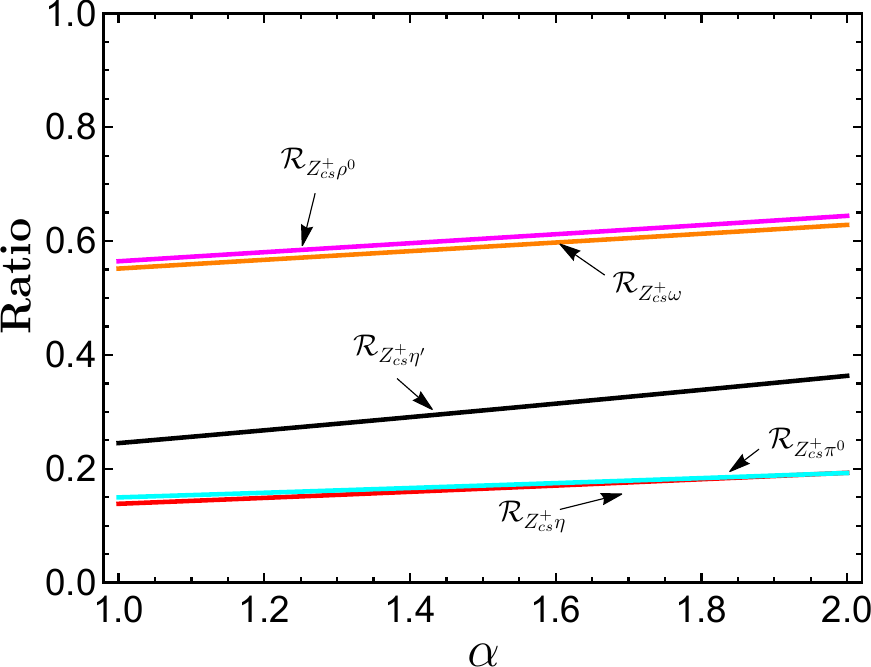}
	\caption{The branching fraction ratios depending on the model parameter $\alpha$. }
	\label{Fig:ratios}
\end{figure}

Now, all the relevant coupling constants and the parameters in the weak transition form factors are prepared except the model parameter $\alpha$ in Eq.~(\ref{Eq:FF}). Since the model parameter $\alpha$ cannot be determined by first-principles methods and it should be of order unity as clarified in Ref.~\cite{Cheng:2004ru}, In the present work, we vary the model parameter $\alpha$ from 1.0 to 2.0 to check the $\alpha$ dependences of the branching fractions of considered processes.

As shown in Fig.~\ref{Fig:br}, the $\alpha$ dependence of branching fractions for considered processes are presented. On the whole, the branching fractions are estimated to around $10^{-4}$ and increase with increasing of $\alpha$. Specifically, the branching fraction of $B^+ \to \phi Z_{cs}^+$ is estimated to be  $(1.75_{-0.50}^{+0.46}) \times 10^{-4}$, where the central value is estimated with $\alpha=1.5$, while the uncertainty comes form the variation of $\alpha$ from 1.0 to 2.0. All the estimated branching fractions are compiled in Table~\ref{Tab:br}, showing $B^+ \to \phi Z_{cs}^+$ as the largest channel, naturally explaining why LHCb first observed $Z_{cs}$ in $B^+ \to J/\psi K^+ \phi$. In Ref.~\cite{Wu:2021ezz}, we investigated the hidden charm decays of the $Z_{cs}$ state, and the branching fraction of $Z_{cs}^+ \to J/\psi K^+$ was estimated to be $(4.0_{-2.7}^{+4.3})\%$. Considering $Z_{cs}$ is a narrow resonance, the branching fraction of the cascade process is
\begin{eqnarray}
	&&\mathcal{B}[B^+ \to  Z_{cs}^+\phi \to J/\psi K^+ \phi ]\nonumber\\
	&&\qquad= \mathcal{B}[B^+ \to  Z_{cs}^+\phi]\times\mathcal{B}[Z_{cs}^+ \to J/\psi K^+]\nonumber\\
	&&\qquad=(7.0_{-5.37}^{+11.3})\times 10^{-6},
\end{eqnarray}
which is comparable with the experimental measurement from the LHCb collaboration shown in Eq.~\eqref{eq:cascade}.

In addition, one can find that the $\alpha$ dependences of the branching fractions are similar. Thus, their ratios should be weakly dependent on the model parameter.  Then we further define the ratios of these branching fractions as,
\begin{eqnarray}\label{eq:R(Zcs+x)}
\mathcal{R}_{Z_{cs}^+ \mathbb{M}}=\frac{\mathcal{B}[B^+\to Z_{cs}^+ \mathbb{M}]}{\mathcal{B}[B^+\to Z_{cs}^+\phi]},
\end{eqnarray}
where $\mathbb{M}$ could be $\rho^0/\omega/\eta^{(\prime)}/\pi^0$. In Fig.~\ref{Fig:ratios}, the $\alpha$ dependence of the ratios are presented, and one can find these ratio are very weakly dependent on the model parameter $\alpha$ as expected. The concrete values of these ratios are collected in  Table.~\ref{Tab:FF}, where the central values are obtained with $\alpha=1.5$, and the uncertainties are resulted from the variation of $\alpha$ from 1.0 to 2.0. From the table one can find the uncertainties of these ratios are very small and these model independent predictions can be tested by further experiment measurements by LHCb and Belle II Collaboration and serve as an important test to the present estimations.

\begin{table}[t]
	\caption{The estimated ratios and fit fractions for the relevant processes in unite of $\%$. The fit fraction of $Z_{cs}^+$ in the $B^+ \to J/\psi K^+ \phi$ are the measured data from the LHCb Collaboration~\cite{LHCb:2021uow}. \label{Tab:FF}}
	\renewcommand\arraystretch{1.5}
	\begin{tabular}{p{2.5cm}<\centering p{2.5cm}<\centering p{2.5cm}<\centering }
		\toprule[1pt]
		Process & Ratio & FF  \\
		\midrule[1pt]
		$B^+\to J/\psi K^+ \phi$  & $\dots$ & $\underline{(9.4\pm2.1\pm3.4)}$ \\
		$B^+\to J/\psi K^+ \omega$  & $0.59_{-0.04}^{+0.04}$ &$0.87_{-0.24}^{+0.24}$\\
		$B^+\to J/\psi K^+ \rho^0$  &$0.60_{-0.03}^{+0.04}$ &  $\dots$\\
		$B^+\to J/\psi K^+ \eta$ &$0.16_{-0.02}^{+0.03}$ & $0.61_{-0.17}^{+0.27}$ \\
		$B^+\to J/\psi K^+ \eta^\prime$ &$0.30_{-0.05}^{+0.06}$ & $4.55_{-1.46}^{+2.22}$ \\
		$B^+\to J/\psi K^+ \pi^0$  &$0.17_{-0.02}^{+0.02}$& $0.07_{-0.01}^{+0.02}$\\
		\bottomrule[1pt]	
	\end{tabular}
\end{table} 

Since most consider decay processes $B^+ \to J/\psi K \mathbb{M}$ have been measured experimentally, thus, we can further consider the fit fraction of the $Z_{cs}^+$ state in the $B^+ \to J/\psi K^+ \mathbb{M}$ decays, which reads,
\begin{eqnarray}\label{eq:ff(Zcs+x)}
	&&FF[B^+\to Z_{cs}^+ \mathbb{M} \to J/\psi K^+ \mathbb{M}] \nonumber\\
	&& \hspace{1cm} =\frac{\mathcal{B}[B^+\to Z_{cs}^+ \mathbb{M} \to J/\psi K^+ \mathbb{M}]}{\mathcal{B}[B^+ \to J/\psi K^+ \mathbb{M}]}.
\end{eqnarray}
With Eqs.\eqref{eq:R(Zcs+x)} and \eqref{eq:ff(Zcs+x)}, the ratios of the fit fraction between two processes could be simplified as,
\begin{eqnarray}
\frac{FF[B^+\to Z_{cs}^+ \mathbb{M} \to J/\psi K^+ \mathbb{M}]}{FF[B^+\to Z_{cs}^+\phi \to J/\psi K^+ \phi]}=\frac{\mathcal{B}[B^+ \to J/\psi  K^+ \phi]}{\mathcal{B}[B^+ \to J/\psi  K^+ \mathbb{M}]}\mathcal{R}_{Z_{cs}^+ \mathbb{M}}.\nonumber\\
\end{eqnarray}
Then, all the fit fraction of $Z_{cs}^+$  in various processes except for $B^+ \to J/\psi K^+ \rho^0$ could be obtained by the measured fit fraction of $Z_{cs}^+$ in the $B^+ \to J/\psi K^+ \phi$ decay, the branching fractions of $B^+ \to J/\psi K^+ \mathbb{M}$ and the  ratio $R_{Z_{cs}^+ \mathbb{M}}$ estimated in the present work, which are also collected in Table~\ref{Tab:FF}. From the table one can find that the fit fraction of $B^+\to Z_{cs}^+\phi \to J/\psi K^+ \phi$ is the largest one, which further explains the results from the LHCb Collaboration that the observation of $Z_{cs}$ in $B^+\to J/\psi K^+ \phi$ process. Moreover, the fit fraction of $Z_{cs}$ in $B^+\to J/\psi K^+\eta^\prime$ process is estimated to be $(4.55_{-1.46}^{+2.22})\%$, which is comparable with the reported fit fraction of $Z_{cs}$ in $B^+ \to J/\psi K^+ \phi$ by the LHCb collaboration, indicating that it is accessible for future experimental measurement.

\section{SUMMARY}
\label{Sec:sum}
In 2021, the BESIII and LHCb Collaboration reported their new observations in the hidden charm and open strange analysis. Two states named $Z_{cs}(3985)$ and $Z_{cs}(4000)$ were seen as the same state since their masses are very close. The mass of the observed state is close to the threshold of $D_s^*D$, leads to a large number of works to investigating them as a molecular state composed of $D_s\bar{D}^*+D_s^*\bar{D}$. Since the large branching ratio of the cascade process $B^+ \to Z_{cs}^+ \phi \to J/\psi K^+ \phi$, on the order of $10^{-6}$, simulating us to  investigate its properties in the $B^+$ decay process. In the present work, we estimated the production of $Z_{cs}^+$ in $B^+$ decay via the triangle loop mechanism. The branching fraction of $B^+ \to Z_{cs}^+\phi \to J/\psi K^+ \phi$ was estimated to be $(7.0_{-5.37}^{+11.3})\times 10^{-6}$, which is comparable with the experimental measurement by the LHCb Collaboration.	

Besides $B^+ \to Z_{cs}^+ \phi$, we also estimated  some other final states such as $Z_{cs}^+ \omega/\rho^0/\eta^{(\prime)}/\pi^0$ in the same mechanism, which are of the order of $10^{-5} \sim 10^{-4}$. Furthermore, with the fit fraction of $Z_{cs}^+$ in $B^+ \to Z_{cs}^+ \phi \to J/\psi K^+ \phi$ reported by LHCb Collaboration, the fit fraction of other processes are also predicted. Our results indicate that the branching fraction of $B^+ \to Z_{cs}^+ \phi$ and the fit fraction of $Z_{cs}$ in $B^+ \to J/\psi K^+ \phi$ are the largest ones in the considered processes, which provides a natural explanation for why the LHCb Collaboration first observed $Z_{cs}^+$ in $B^+ \to J/\psi K^+ \phi$. The fit fraction of $Z_{cs}^+$ in $B^+ \to Z_{cs}^+ \eta^\prime \to J/\psi K^+ \eta^\prime$ is estimated to be $(4.55_{-1.46}^{+2.22})\%$, which is also comparable with the experimental measurement by the LHCb Collaboration. Thus, we suggest searching for $Z_{cs}^+$ in the $B^+ \to J/\psi K^+ \eta^\prime$ in future experiments.

\section*{ACKNOWLEDGMENTS}
 This work is supported by the National Natural Science Foundation of China under the Grant Nos. 12175037, 12335001 and 12405093, as well as supported, in part, by National Key Research and Development Program under Grant No.2024YFA1610504.

\bibliographystyle{unsrt}
\bibliography{references.bib}
\end{document}